\documentclass{elsarticle}
\usepackage{amsmath}
\usepackage{textcomp}
\usepackage{multirow}

\DeclareGraphicsExtensions{.pdf,.png,.jpg} 

\usepackage[
pdfauthor={Olivier Bourrion},
pdftitle={Development of a multifuntion module for the neutron electric dipole moment at PSI},
pdfsubject={nEDM},
pdfkeywords={},
colorlinks={true},
linkcolor=red,
citecolor=green,
urlcolor=cyan,
pdfstartview={FitH}
]{hyperref}

\newcommand{\fig}{fig.~}
\begin{document}
\begin{frontmatter}

\title{Development of a multifunction module for the neutron electric dipole moment experiment at PSI}

\author[LPSC]{O.~Bourrion\corref{cor1}}
\ead{olivier.bourrion@lpsc.in2p3.fr}
\author[LPSC]{G.~Pignol}
\author[LPSC]{D.~Rebreyend}
\author[LPSC]{C.~Vescovi}

\cortext[cor1]{Corresponding author}
\address[LPSC]{Laboratoire de Physique Subatomique et de Cosmologie,\\ 
Universit\'e Joseph Fourier Grenoble 1,\\
  CNRS/IN2P3, Institut Polytechnique de Grenoble,\\
  53, rue des Martyrs, Grenoble, France}

\begin{abstract}
Experiments aiming at measuring the neutron electric dipole moment (nEDM) are at the forefront of precision measurements and demand instrumentation of increasing sensitivity and reliability. 
In this paper, we report on the development of  a dedicated acquisition and control electronics board for the nEDM experiment at the Paul Scherrer Institute (PSI) in Switzerland. 
This multifunction module is based on a FPGA (Field-programmable gate array) which allows an optimal combination of versatility and evolution capacities. 
\end{abstract}

\end{frontmatter}

\section{Introduction}
\label{introduction}

The search for a non-zero electric dipole moment (EDM) of the neutron, initiated 60 years ago by Norman Ramsey\cite{Ramsey}, 
continues today to motivate experimental activity with increasing precision. 
A non-zero electric dipole moment for a spin 1/2 particle, such as the neutron, implies the violation of both parity (P) and time reversal (T) symmetries; according to the CPT theorem this would also imply a violation of the combined CP symmetry.  
Since CP violation beyond the standard model of particle physics is required to explain the asymmetry between matter and antimatter, the so-called baryon asymmetry of our Universe, 
 electric dipole moments provide a window on the early Universe when the baryon asymmetry was generated. 
For a recent review on the connection between neutron physics and cosmology, see \cite{Dubbers} and references therein.

The best measurement of the neutron EDM \cite{Baker} was obtained at the Institute Laue-Langevin (ILL) in Grenoble, France and provided the upper limit $\rm |d_n|< 2.9 \times 10^{-26} e\cdot cm$ (90\% C.L.). It was extracted from the precession frequency of trapped ultra-cold neutrons (UCN) in a weak ($\rm B_0$= 1\,\textmu T) magnetic field, using Ramsey's method of separated oscillating fields\cite{RamseyMethod}.
In addition, the apparatus has also been used to perform sensitive tests of Lorentz invariance with neutrons \cite{Lorentz1,Lorentz2,Lorentz3}, where a  daily modulation of the Larmor frequency was searched for.

The new experiment, installed at the Paul Scherrer Institute (PSI), Villigen, Switzerland, uses an upgraded version of this apparatus\cite{PSIEDM} and aims at improving the sensitivity by an order of magnitude, thanks to a higher UCN density.
The development of a completely renewed data-acquisition system (DAQ) was part of the numerous upgrades.
To this end, new electronic modules were needed to execute all requested actions and control the experiment. Instead of multiple specialized modules, we have opted for a central and multifunction module. With such a solution, we expect to benefit from the limited number of connectors, thus minimizing the delicate problem of bad contacts. 

After a brief summary of the experiment principle, a description of a typical measurement cycle will be given in section~\ref{part2}.
The experiment most stringent requirements are given in section~\ref{requirements}. 
Section~\ref{hardwareSec} will present in details the electronic board, and section~\ref{firmwareSec} the embedded firmware. Finally, section~\ref{protoValSec} will report on some of the achieved performances.

\section{Description of the measurement sequence}
\label{part2}

\begin{figure}[ht]
\begin{center}
\includegraphics[angle=-90,width=0.95\textwidth]{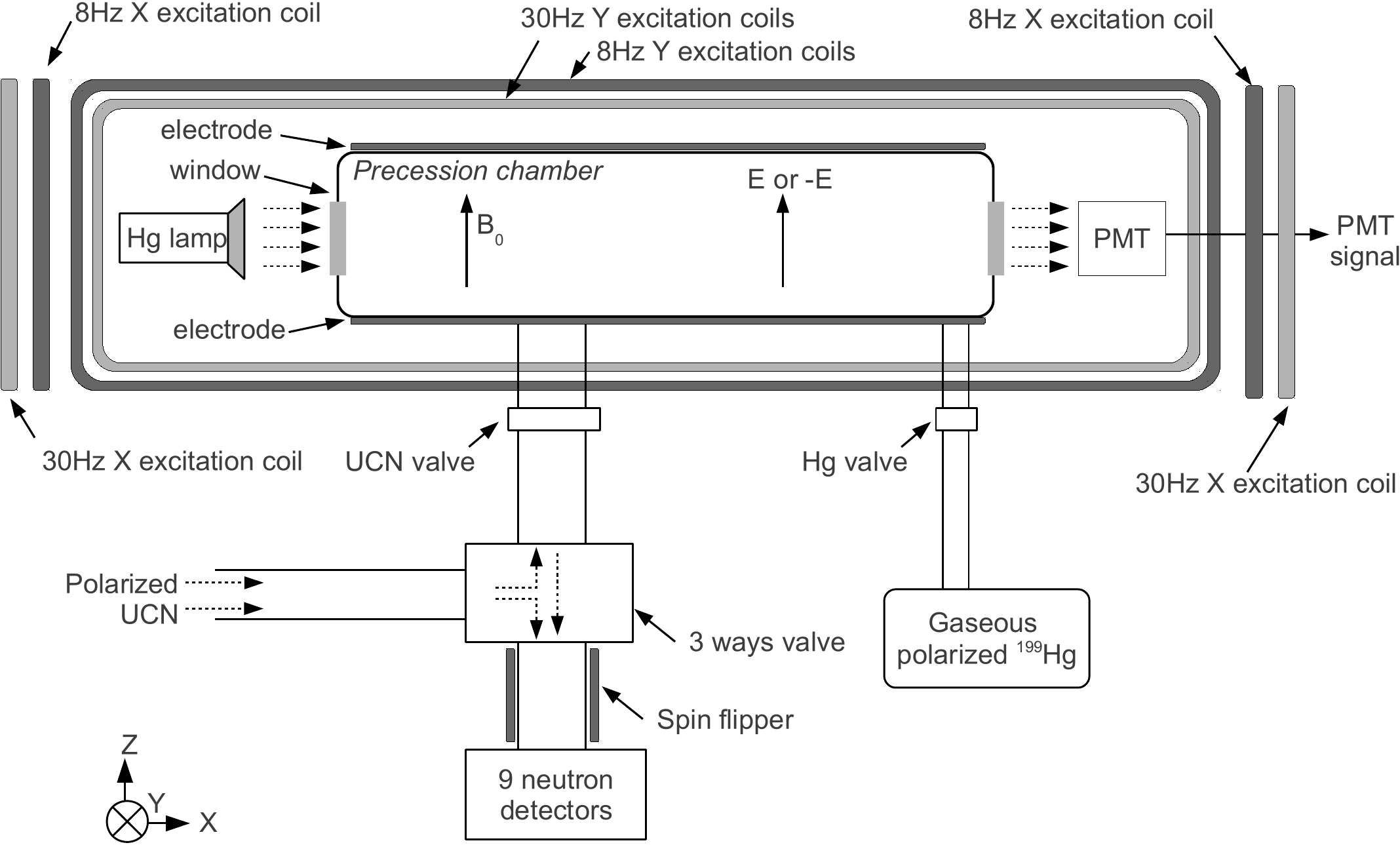}
\caption{Schematic of the experiment.}
\label{ExperimentDesc} 
\end{center}
\end{figure}

The PSI EDM experiment\cite{PSIEDM} uses the RAL/SUSSEX/ILL apparatus, connected to the newly built PSI UCN source\cite{UCNsource}. The main elements of the apparatus are depicted in fig.~\ref{ExperimentDesc}.  
The central part, the precession chamber, is a 21 liters cylindrical bottle (D=47\,cm, H= 12\,cm) used to store polarized ultra-cold neutrons and polarized $^{199}$Hg atoms. 
This chamber is shielded against the ambient magnetic field with a four-layer mu-metal assembly. 
A highly homogeneous vertical magnetic field $\rm B_0$= 1\,\textmu T is generated inside the mu-metal shield by a set of coils. 
Additionally, oscillating transverse (to $\rm B_0$) fields can be applied by pairs of coils in the X or Y directions (\fig\ref{ExperimentDesc}).
The bottom and top parts of the precession chamber also serve as electrodes to generate a strong electric field (typically $120 \, {\rm kV} / 12 \, {\rm cm}$), either parallel or antiparallel to the $\rm B_0$ magnetic field. 

The principle of the experiment consists in measuring the neutrons' Larmor precession frequency using Ramsey's method of separated oscillating fields. 
The neutron Larmor frequency is proportional to the magnitude of $\rm B_0$, plus possibly a small frequency shift due to the neutron EDM. 
A frequency difference, proportional to E, between the parallel and antiparallel configurations would be the signature for a non-zero EDM.
However, the small drifts of the $\rm B_0$ field during the experiment could easily hide the signal. 
To measure and correct for these fluctuations, an atomic magnetometer using a vapor of polarized $^{199}$Hg atoms cohabiting in the same volume as the UCN is used\cite{Green}.

\begin{figure}[ht]
\begin{center}
\includegraphics[angle=-90,width=0.8\textwidth]{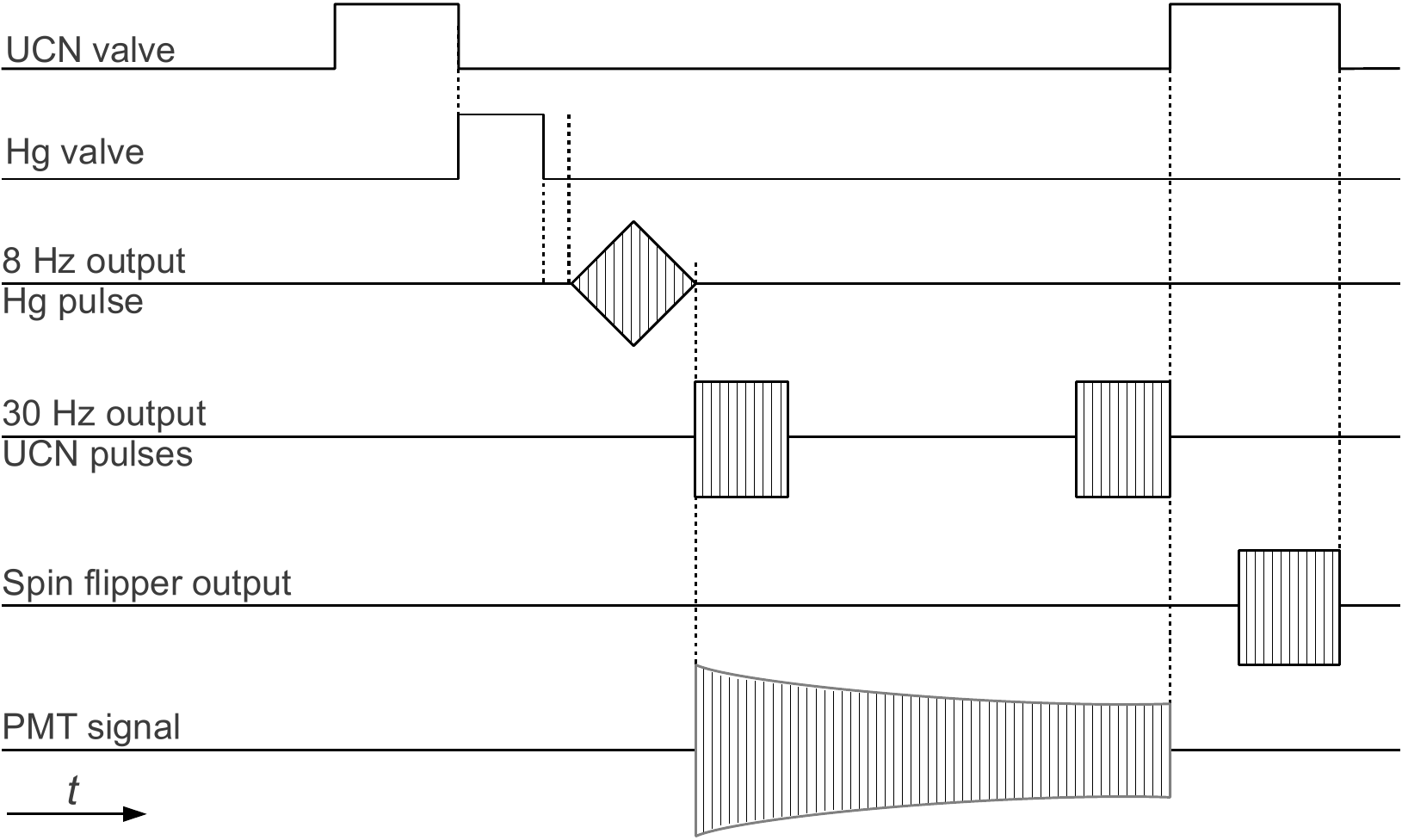}
\caption{Snapshot of standard EDM cycle.}
\label{standardEdmCycle} 
\end{center}
\end{figure}

A standard EDM cycle, sketched in \fig\ref{standardEdmCycle}, is composed of seven successive steps detailed below. 

\begin{enumerate}
\item {\bf UCN fill:}
 Polarized UCN are guided toward the storage chamber. The filling time, controlled by the UCN valve, lasts  typically 20\,s. 

 \item {\bf Mercury fill:} 
The mercury valve is opened for 2 s, to let the $^{199}$Hg gas, optically pumped in the polarization cell, diffuse in the precession chamber.

\item {\bf Mercury pulse:}
A rotating transverse magnetic field is generated for 2\,s by two pairs of coils. The signal frequency is equal to the Larmor frequency of the $^{199}$Hg atoms (approx. 7.6\,Hz for $\rm B_0$=1\,\textmu T), measured during the previous cycle.
The signals feeding the X and Y coils are in phase quadrature sine waves and their amplitudes are adjusted so that the Hg spins rotate by $\pi/2$\,rad. A triangular envelope was chosen in order to minimize the effect of the mercury pulse on the neutrons spins.

\item {\bf First UCN pulse:} 
Similarly to the mercury pulse, two pairs of coils induce a $\pi/2$ flip of the neutrons spins. The signal frequency is also calculated from the mercury Larmor frequency measured at the previous cycle. 
However, for the neutrons, the excitation frequency is changed, from one cycle to the next, using two working points, on both sides of the central Ramsey resonance fringe.
These four working points are used to fit the Larmor neutron precession frequency (approx. 29\,Hz for $\rm B_0$=1\,\textmu T). 

\item {\bf Free precession:} 
For a typical duration of 200\,s, both neutron and mercury spins precess in the horizontal plane around $\rm B_0$. 
The mercury free precession is optically monitored, using through-going polarized light produced by a mercury lamp. 
The light intensity, measured with a photomultiplier (PMT), is modulated at the Larmor frequency of the mercury spins. 
The PMT signal is sampled with the DAQ electronics and an estimate of the frequency is extracted. As discussed above,
the estimated frequency $f_{\rm Hg}$ serves to set the frequency of both  mercury and neutron pulses at the next cycle, thus accounting for the drifts in magnetic field. 
\item {\bf Second UCN pulse:} 
At the end of the precession time, a second neutron $\pi/2$ pulse, in phase with the first one, is generated, thus completing the Ramsey sequence. 
\item {\bf Neutron polarization measurement:} 
The UCN valve is opened to empty the precession chamber and neutrons fall down toward a neutron detector. 
On their way, neutrons encounter a magnetized ferromagnetic foil that serves as spin analyzer by letting only one spin component go through. 
After a fixed amount time, an Adiabatic Fast Passage (AFP) spin flipper is turned on to count the number of neutrons in the other component. From this information and the knowledge of the $\pi/2$ pulse frequency, the neutron Larmor frequency can be extracted.

\end{enumerate}

The described sequence is repeated continuously to accumulate statistics, with periodic reversal of the electric field. 
Two years of data-taking at the PSI UCN source are planned to record about 50,000 cycles and gain an order of magnitude in sensitivity.

\section{Summary of experimental requirements}
\label{requirements}
In this section, the most stringent experimental requirements are discussed.

\begin{itemize}

\item Experiment timing resolution\\
The sequence of actions described above must be controlled with high precision.
The most demanding actions correspond to the control of the $\pi$/2 pulses and of the neutron counting sequence.
An error on the  $\pi$/2 pulse duration will affect the precision of the frequency extraction but not the frequency itself.
For the mercury as well as for the neutron  $\pi$/2 pulses, we have estimated that a 1\,ms deviation from a 2\,s pulse would result in a precision loss of 10$^{-6}$ relative to the extracted frequency, a negligible amount.
Likewise, having a resolution of $\pm$1\,ms on the counting time sequence would induce an error on the UP/DOWN asymmetry much smaller than the statistical fluctuations, which is sufficient.

\item Wave generator frequency resolution\\
The neutron Larmor precession frequency is extracted from the combination of the applied frequency of the $\pi$/2 pulse and of the neutron counting.
The resolution of the wave generator should be therefore significantly better than the expected statistical precision $\sigma f_n/f_n \approx 3 \times 10^{-6}$, i.e at 30\,Hz a resolution of about 0.1\,mHz is required.

\item Scaler counting rate\\
With the increased neutron density provided by the new PSI source, peak counting rates as high as 10$^6$\,Hz can be reached.

\item Mercury signal encoding resolution\\
The precision on the extracted mercury Larmor frequency is directly proportional to the signal-to-noise ratio (SNR) of the precession signal measurement.
Our current SNR is typically 2000 at the beginning of the precession, which limits us to a frequency precision of 0.5\,\textmu Hz.  
In the future, we plan to reach a SNR of about 10000 to approach a 0.1\,\textmu Hz precision. 
Therefore to have a significant margin over the quantization noise, a 16 bit ADC is required.

\end{itemize}


\section{Hardware description}
\label{hardwareSec}

\begin{figure}[ht]
\begin{center}
\includegraphics[angle=0,width=0.95\textwidth]{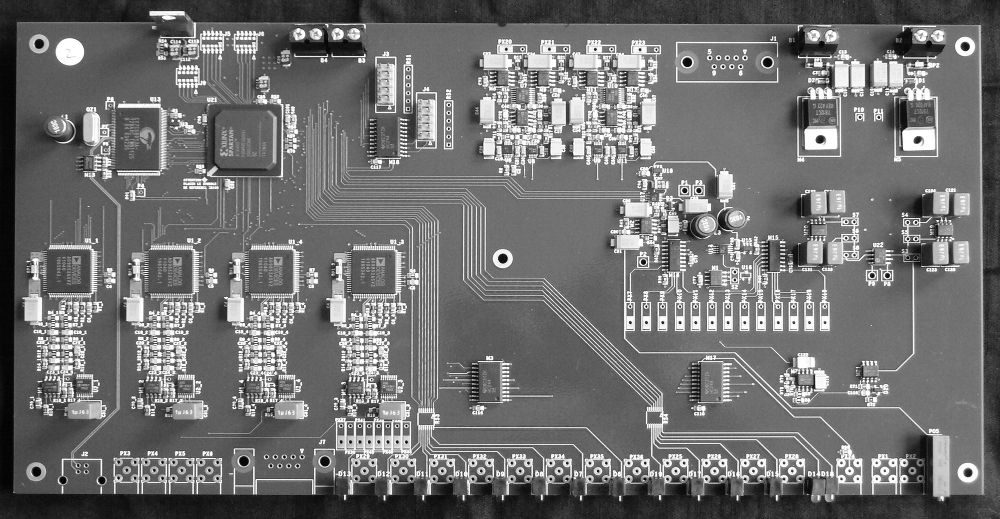}
\caption{nEDM experiment acquisition and control electronic board picture. The board has dimensions of $\rm 300\,mm \times 150\,mm$.}
\label{boardPicture} 
\end{center}
\end{figure}

\begin{figure}[ht]
\begin{center}
\includegraphics[angle=-90,width=0.95\textwidth]{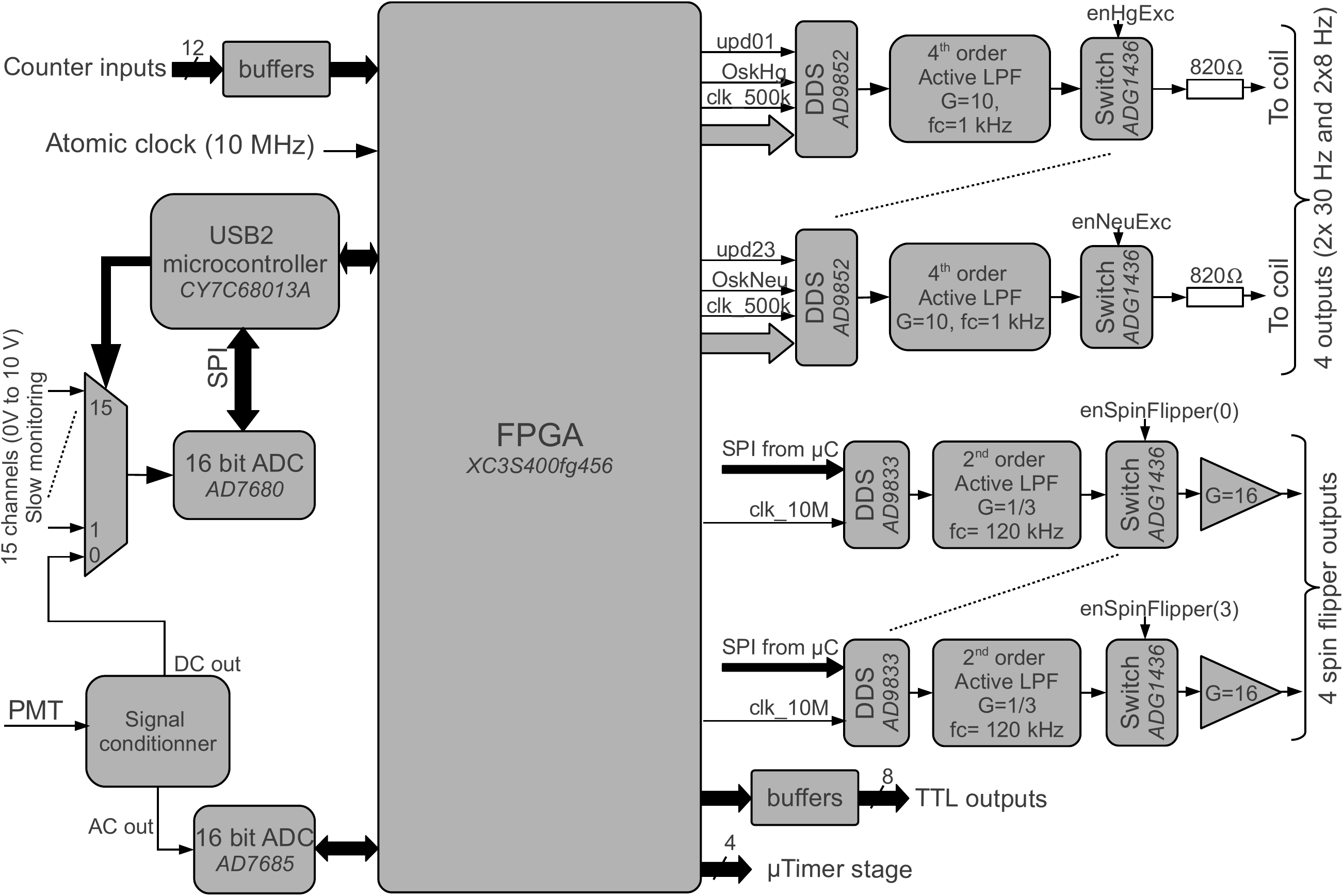}
\caption{Block diagram of the electronic board.}
\label{BlocCarteV2} 
\end{center}
\end{figure}

A picture of the electronic board is shown in \fig\ref{boardPicture} and its block diagram is shown in \fig\ref{BlocCarteV2}. 
It is designed around a Field Programmable Gate Array (FPGA) (Xilinx XC3S400fg456), whose main purposes are to perform the system precise operation sequencing (`micro-timer') and the data acquisition. 
The communication with the board is done via a USB2.0 capable micro-controller (Cypress CY7C68013A), this allows to perform the system control and the data readout.
Furthermore, it is used to load the FPGA firmware at board power up and to perform the monitoring of 16 input channels with a 16 bit ADC. 
It directly controls the signal multiplexer and reads out the ADC.
Channel 0 is reserved for reading the Photo Multiplier Tube (PMT) DC signal.

A Rb atomic clock\footnote{GPSReference-2000 from spectra time, short term stability: 3$\times 10^{-12}$/10\,s and long term stability: $<$ 3$\times 10^{-11}$/month.}, operating at 10\,MHz, is directly connected to the FPGA in order to reference the whole system (signal generation and measurements).
The various sine wave generation is ensured by two kinds of Direct Digital Synthesis (DDS) circuits whose outputs are connected or disconnected by FPGA driven switches.
The first DDS kind (AD9833) is used to provide the spin flipper signals (typical frequency of 19\,kHz). 
They have a 28 bit resolution phase accumulator and thus can achieve a frequency resolution of 37\,mHz when supplied with a 10\,MHz reference clock. 
These DDS are configured by the micro-controller via a Serial Peripheral Interface (SPI) link. 
Each DDS output signal, which has an amplitude resolution of 10 bit, is filtered by an active second order Low Pass Filter (LPF) before passing through the FPGA controlled switch. 
The switch output is then amplified by 16 before being provided to the spin flipper coils.

\begin{figure}[ht]
\begin{center}
\includegraphics[angle=0,width=0.6\textwidth]{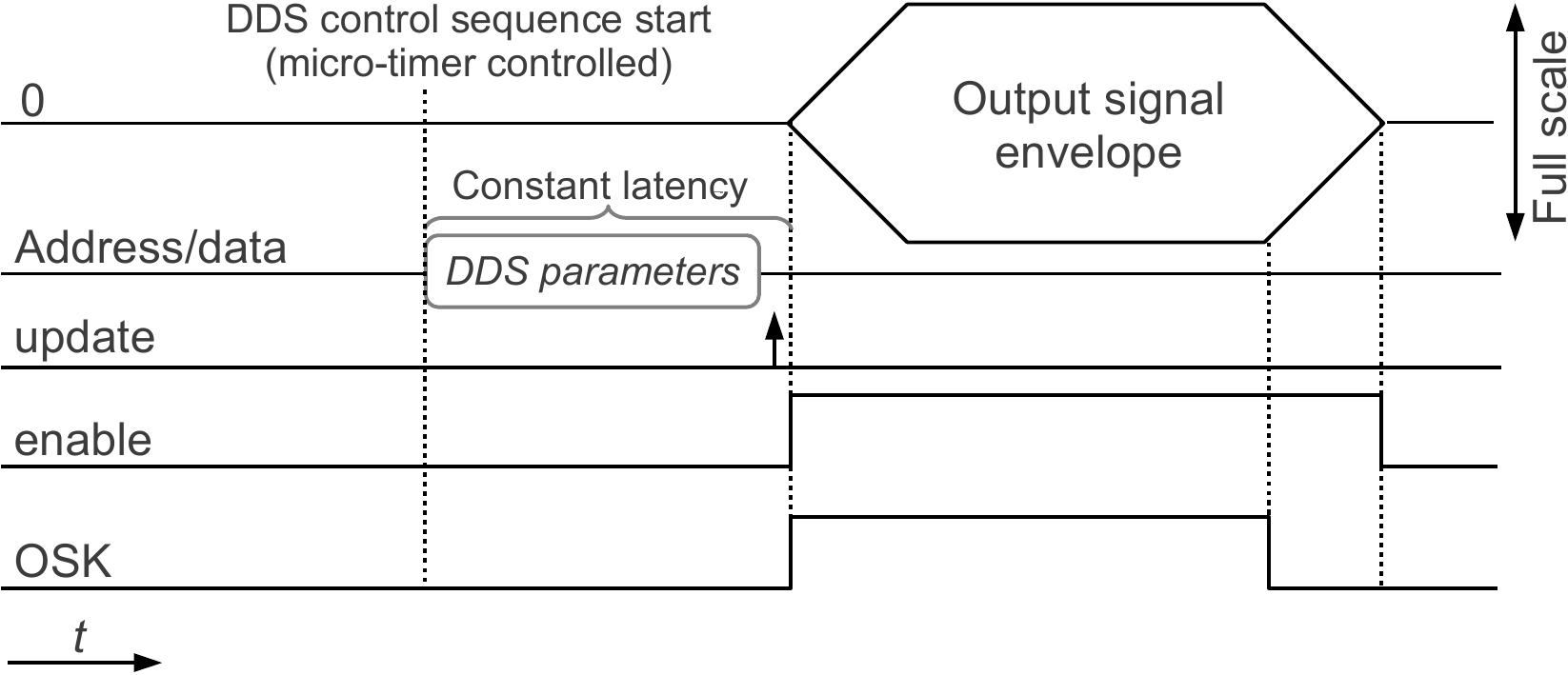}
\caption{Example of a full control sequence allowing to adjust the start phase, ramping and switch control with a deterministic timing.}
\label{OSKsequence} 
\end{center}
\end{figure}

The second DDS kind (AD9852) is used to provide the two neutron (around 30\,Hz) and two Mercury (around 8\,Hz) excitation signals with a very high precision in frequency.
They feature a 48 bit phase accumulator and, given the fact that they are referenced at 500\,kHz (a divided version of the atomic clock), they have a tuning resolution of 1.77\,nHz.
This resolution which is therefore far better than the experimental requirement of 0.1\,mHz. 
Also these DDS offer the possibility to add an offset to the internal phase accumulators result. Consequently, it is possible to generate the excitation signal pair with a known phase relationship by updating them synchronously with the same phase increment value but with different phase offset. This synchronous update is achieved the FPGA which directly controls both DDS update input pin of a pair. Moreover, by performing this operation slightly before closing the DDS output switches, which are also FPGA controlled, it is possible to adjust the start phases of both sine waves (see \fig\ref{OSKsequence}).

\begin{figure}[ht]
\begin{center}
\includegraphics[angle=0,width=0.6\textwidth]{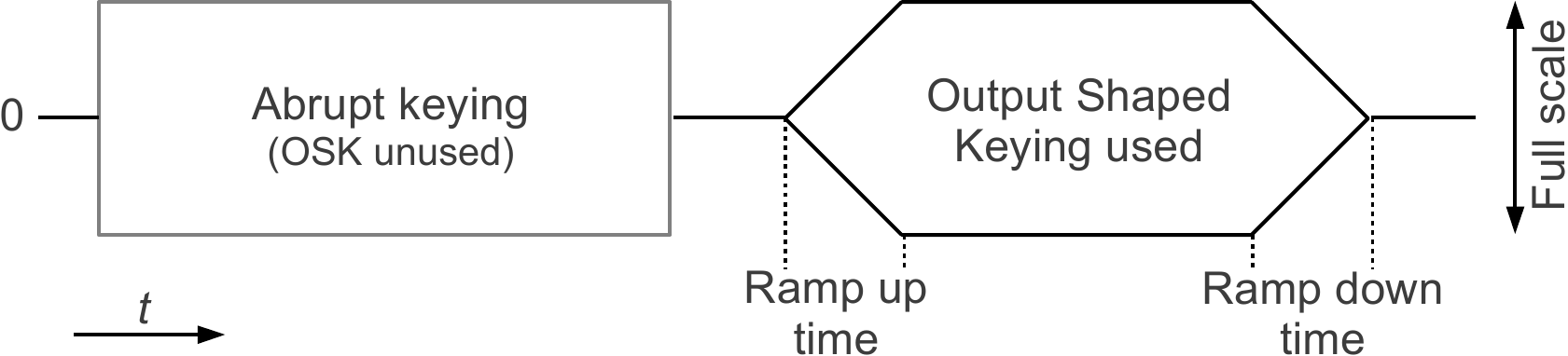}
\caption{Sketch of the output amplitude as a function of time for the abrupt and OSK modes. While the abrupt mode toggle the output amplitude instantly to full scale, the OSK feature allow the control the sine/cosine amplitude as a function of the time.}
\label{OFSKPrinciple} 
\end{center}
\end{figure}

Another advantage, is that the sine wave output amplitude, which has an amplitude resolution of 12 bit, can be digitally adjusted either by a constant factor or by using the Output Shape Keying (OSK) feature. 
As illustrated in \fig\ref{OFSKPrinciple}, this allows the control of the signal amplitude as a function of time and therefore  
to generate the required triangular shape envelope instead of the abrupt rectangular shape.
With the DDS clock running at 500\,kHz, the ramping time can be adjusted from 32.768\,ms up to 2.096\,s with a step resolution of 8.192\,ms. 
The ramping (up or down) is deterministically activated by the micro-timer embedded in the FPGA via a dedicated pin (labeled OSK). 
An example of the control sequence allowing to adjust the initial phase, the ramping and the switch control is displayed in \fig\ref{OSKsequence}.
Each DDS output signal is amplified with an adjustable gain up to a factor ten and filtered by an active fourth order Low Pass Filter (LPF) before passing through the FPGA controlled switch. 
A serial resistor (820$\Omega$) has been added at the circuit output in order to have a current-source-like output. 
Consequently, the magnetic field generated is less sensitive to the coil resistance variation due to thermal fluctuations (typical coil resistance is around 10\,$\Omega$).

\begin{figure}[ht]
\begin{center}
\includegraphics[angle=0,width=0.6\textwidth]{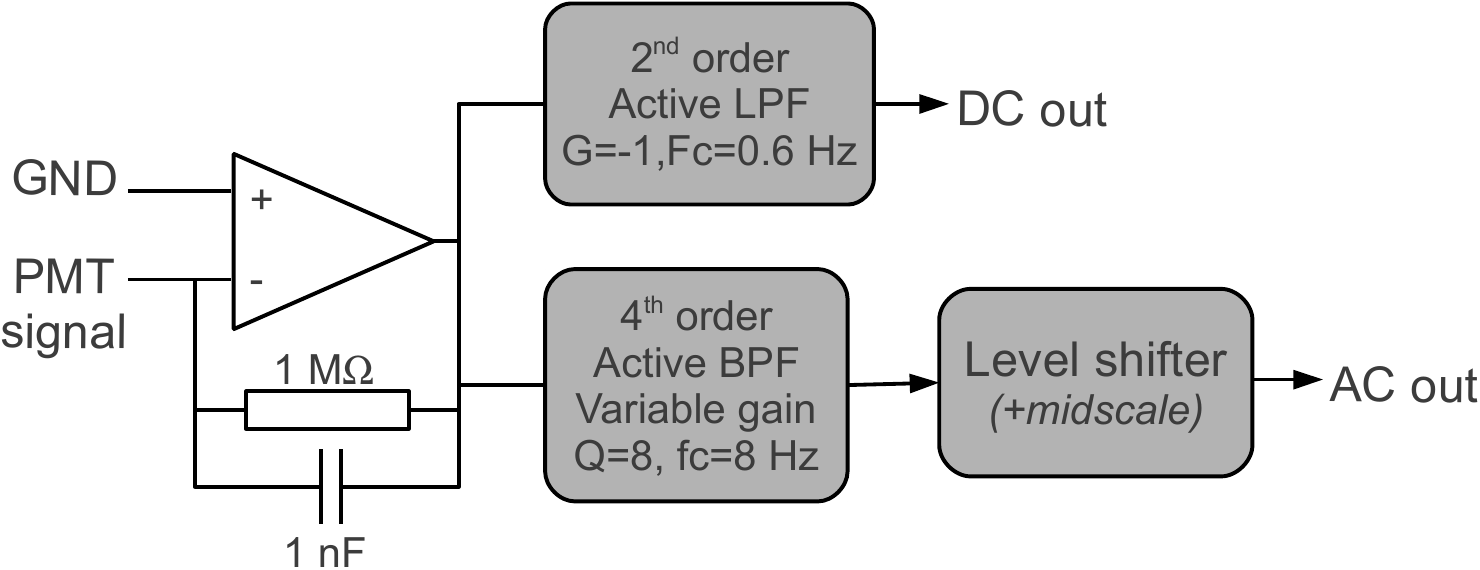}
\caption{Block diagram of the PMT signal conditioning. This block converts the current signal in a voltage signal and separate the DC and AC parts of the signal.}
\label{PMTFE} 
\end{center}
\end{figure}

The PMT signal is connected to a conditioning block detailed in \fig\ref{PMTFE}. 
The first stage converts the current signal in a voltage signal with a current preamplifier having a 1\,M$\Omega$ resistor feedback.
The alternating (AC) and continuous (DC) components of the signal are then separated: the DC component is extracted via a second order low pass filter while the AC part passes through a Q\,=\,8 band pass filter centered around 8\,Hz. 
 The resulting signal is level shifted by half of the ADC full-scale in order to fit in the available ADC range.

Some additional features of this module include:
\begin{itemize}
\item the possibility to control eight external electromechanical devices synchronously to the micro-timer with the provided buffered TTL outputs. 
\item the generation of a real-time four bit signal at each micro-timer step, encoding the step number to permit the synchronization with other electronic modules.
\item the availability of twelve 32 bit scalers to record the neutron counts at the end of a EDM cycle.
\end{itemize}
\section{Firmware description}
\label{firmwareSec}
\begin{figure}[ht]
\begin{center}
\includegraphics[angle=0,width=0.95\textwidth]{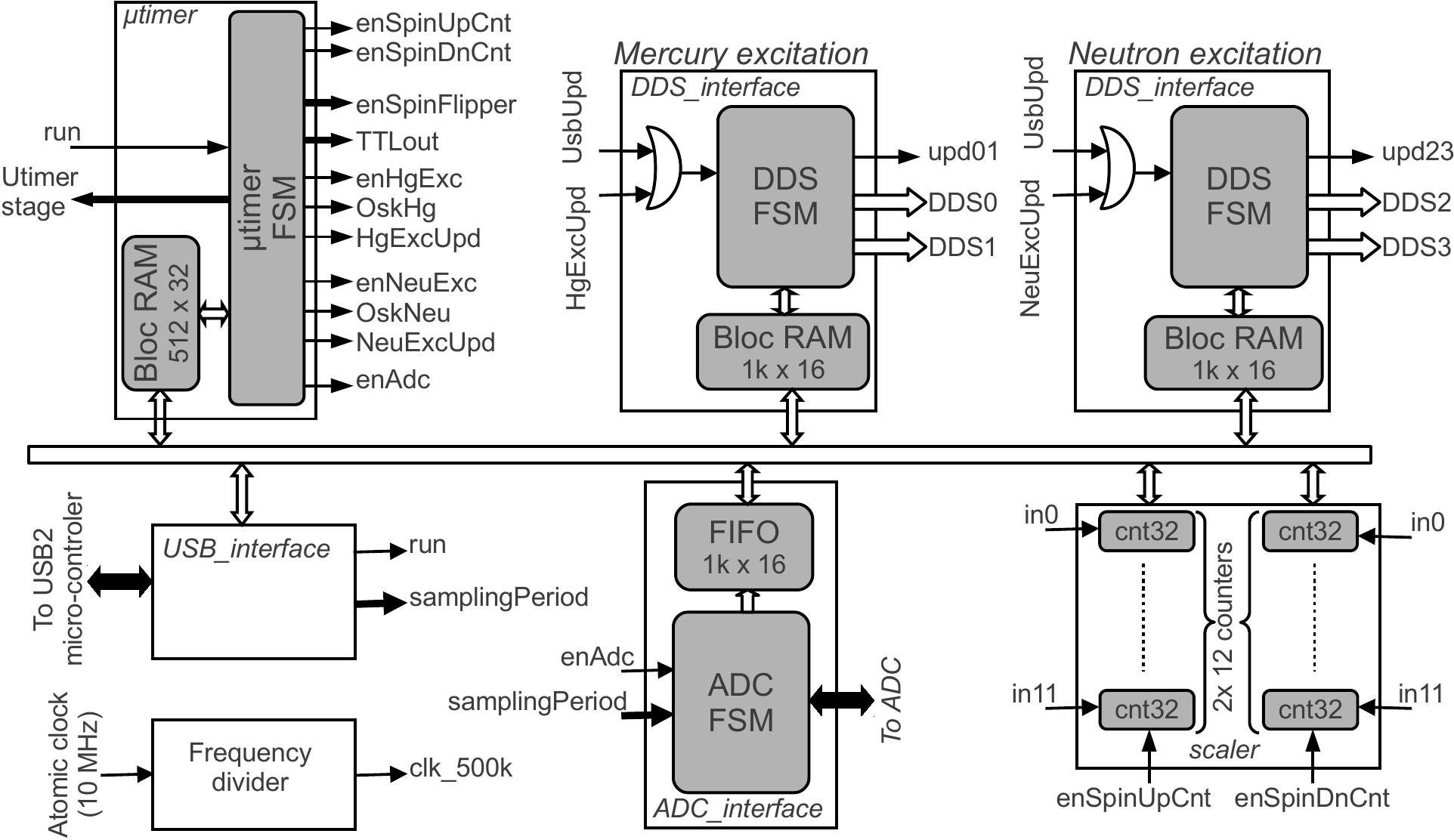}
\caption{Overview of the firmware. It is composed of seven different blocks: the frequency divider, the USB interface for communication, the scalers, the micro-timer for precise operation sequencing, the ADC interface for data acquisition and the two DDS interface blocks.}
\label{BlocFPGAV2} 
\end{center}
\end{figure}

An overview of the FPGA firmware which is the core of the system and whose main purposes are to perform the system precise operation sequencing and the data acquisition is shown in \fig\ref{BlocFPGAV2}. It is composed of seven different blocks: the frequency divider, the USB interface for communication, the scalers, the micro-timer for precise operation sequencing the ADC interface for data acquisition and the two DDS interface blocks.

The frequency divider is used to divide the atomic clock frequency by a factor 20 for referencing of the DDS. As explained in section~\ref{hardwareSec}, the frequency must be lowered in order to reach the highest resolution on the output signal frequency but also to be able to have slow ramping amplitudes. This division is performed by a 0 to 9 counter which keeps a 50\% duty cycle.
 
The micro-timer block is designed to manage up to 16 time sequences. 
Each sequence has a programmable duration, that can be adjusted with a 1\,\textmu s step resolution in a 1 to 2\textsuperscript{32}-1 range, and can manage up to 32 different actions. 
The actions to be executed during each step are controlled by a 32-bit word (action mask), each action having a dedicated bit.
The 32 micro-timer parameters, the 16 durations and the 16 action masks, are written in the dual port block RAM by the USB interface before starting the time sequence execution.
The micro-timer block is managed by a simple two states Finite State Machine (FSM) which starts the time sequence execution when the run signal, provided by the USB interface, is activated.
Before executing each instruction cycle, the corresponding parameters are read from the memory.
Additionally, the 4 Most Significant Bit (MSB) of the memory read pointer are used to reflect the current sequence in progress. 
This information is available for USB readout and on the board for external synchronization. 
The possible micro-timer actions are listed below:
\begin{itemize}
 \item The eight `TTLout' may activate external electro-mechanical actuators (UCN and Hg valves).
 \item `HgExcUpd' and `NeuExcUpd': signals to trigger the DDS parameters update for mercury and neutron, respectively.
 \item `enHgExc' and `enNeuExc': enabling of signals generation for mercury and neutron respectively.
 \item `OskHg' and `OskNeu': initialization of the ramping of the sine wave amplitude (see figure~\ref{OFSKPrinciple}).
 \item `enAdc': enabling of PMT signal acquisition.
 \item The four  `enSpinFlipper' are used to activate the selected 19\,kHz sine wave generator.
 \item The `enSpinUpCnt' and `enSpinDnCnt' bit are respectively used to allow the 12 associated scalers to count during spin up and spin down.
\end{itemize}

The DDS interface blocks are used to simultaneously configure the DDS pairs (mercury or neutron) with the parameters provided by the USB interface in their embedded block memory. 
Each DDS requires up to 40 configuration bytes. Consequently, the two DDS parameters bytes are concatenated in a single word and the whole configuration is conveniently stored in the 40 first memory words (16 bit).
The FSM, which manages the communication protocol with the DDS, initiates the parameter transfer upon reception of the update signal provided by the micro-timer table or by the USB interface.

The firmware features a scaler containing two banks of twelve 32-bit counters, each counter having the capacity to count with a frequency up to 100\,MHz. These counters are operated in a gated mode with the gating signals provided by the micro-timer as previously described.

The PMT signal acquisition is managed by the ADC interface block which performs the data acquisition at the selected sampling period and when activated by the micro-timer. 
The digitized data are written in the embedded output FIFO and real-time data-readout is realized via the USB interface. 
The sampling period, which is a sub-division of the atomic reference clock,  can be adjusted by steps of 100\,ns between a minimum of 10\,\textmu s and a maximum of 1677\,ms. 
It should be noted that running at a conversion rate significantly higher than the signal frequency (8\,Hz) can be an asset. Indeed, at the price of more computing power, it is possible to apply an off-line sharp digital filtering to further reduce the bandwidth given by the analog band pass filter. Hence, the signal-to-noise ratio can be further improved.

\section{Board performances}
\label{protoValSec}

\begin{figure}[ht]
\begin{center}
\includegraphics[angle=0,width=0.95\textwidth]{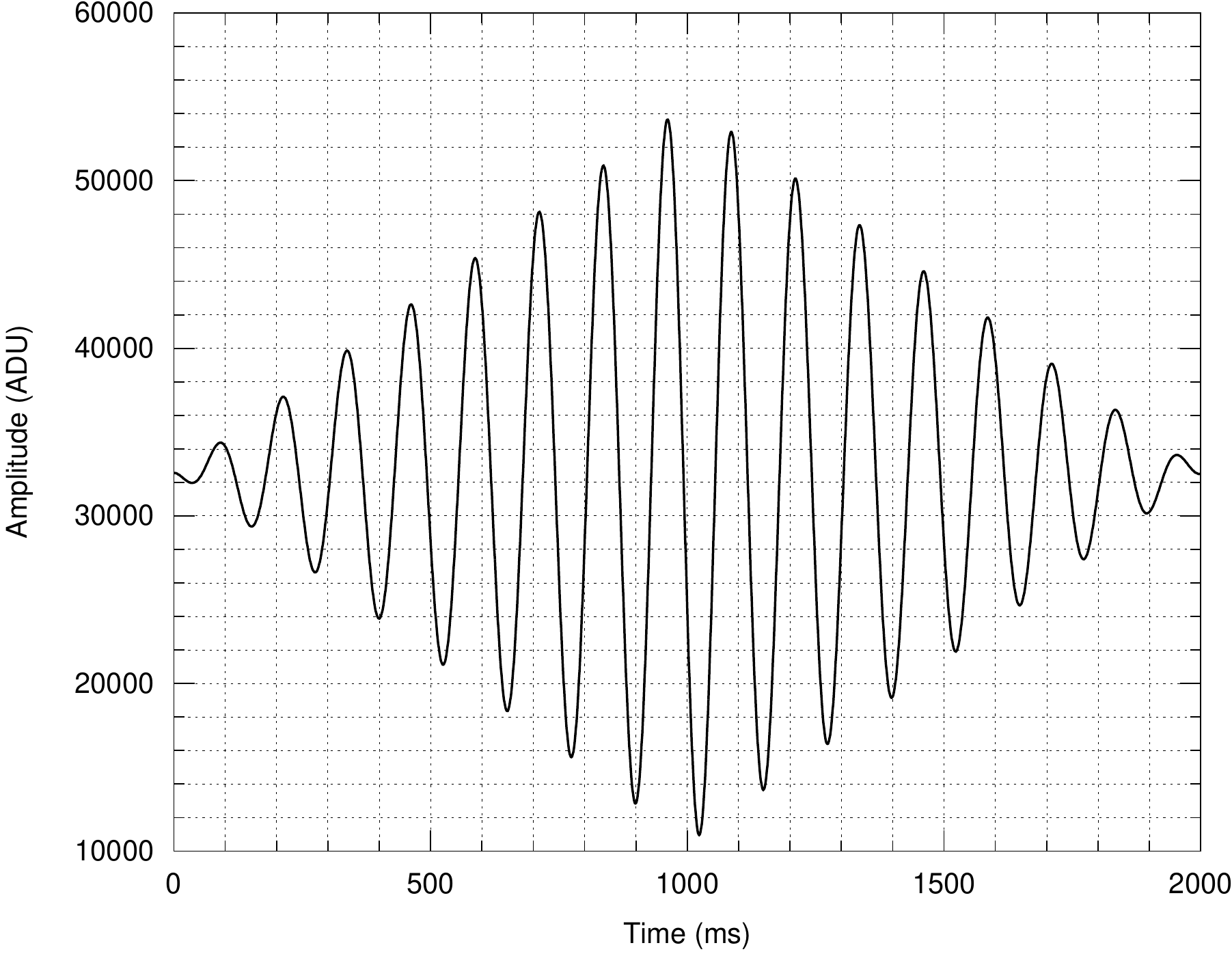}
\caption{Capture of an 8\,Hz excitation sine wave, having a duration of 2\,s and 1\,s rise and fall time. The signal was sampled at 1\,kHz with the boards own ADC.}
\label{triangle-2s_1kHz} 
\end{center}
\end{figure}

A series of tests was conducted to check the main performances of the board and ensure it fulfills the requirements.
As an example of the various basic functional tests, we show in \fig\ref{triangle-2s_1kHz} the generation of a sine wave signal with a triangular envelope. This 8\,Hz sine wave having a duration of 2\,s and 1\,s rise and fall times, was sampled at 1\,kHz with the board's own ADC.


We have quantified the quality of the waveform by fitting the signal with phase, starting time and maximum amplitude as free parameters (the frequency and the total duration of the signal were fixed to preset values). We found a residual better than 10\textsuperscript{-3} relative to the maximum amplitude.

The most critical and demanding function of the board concerns the PMT signal treatment. Indeed, as discussed in section~\ref{part2}, the Hg co-magnetometer is a key element of the nEDM apparatus since, by measuring the mercury precession frequency, we can precisely measure the magnetic field seen by the neutrons and correct for its fluctuations. Systematic tests were therefore carried out to study the measurement chain performances.

\begin{figure}[ht]
\begin{center}
\includegraphics[angle=-0,width=0.95\textwidth]{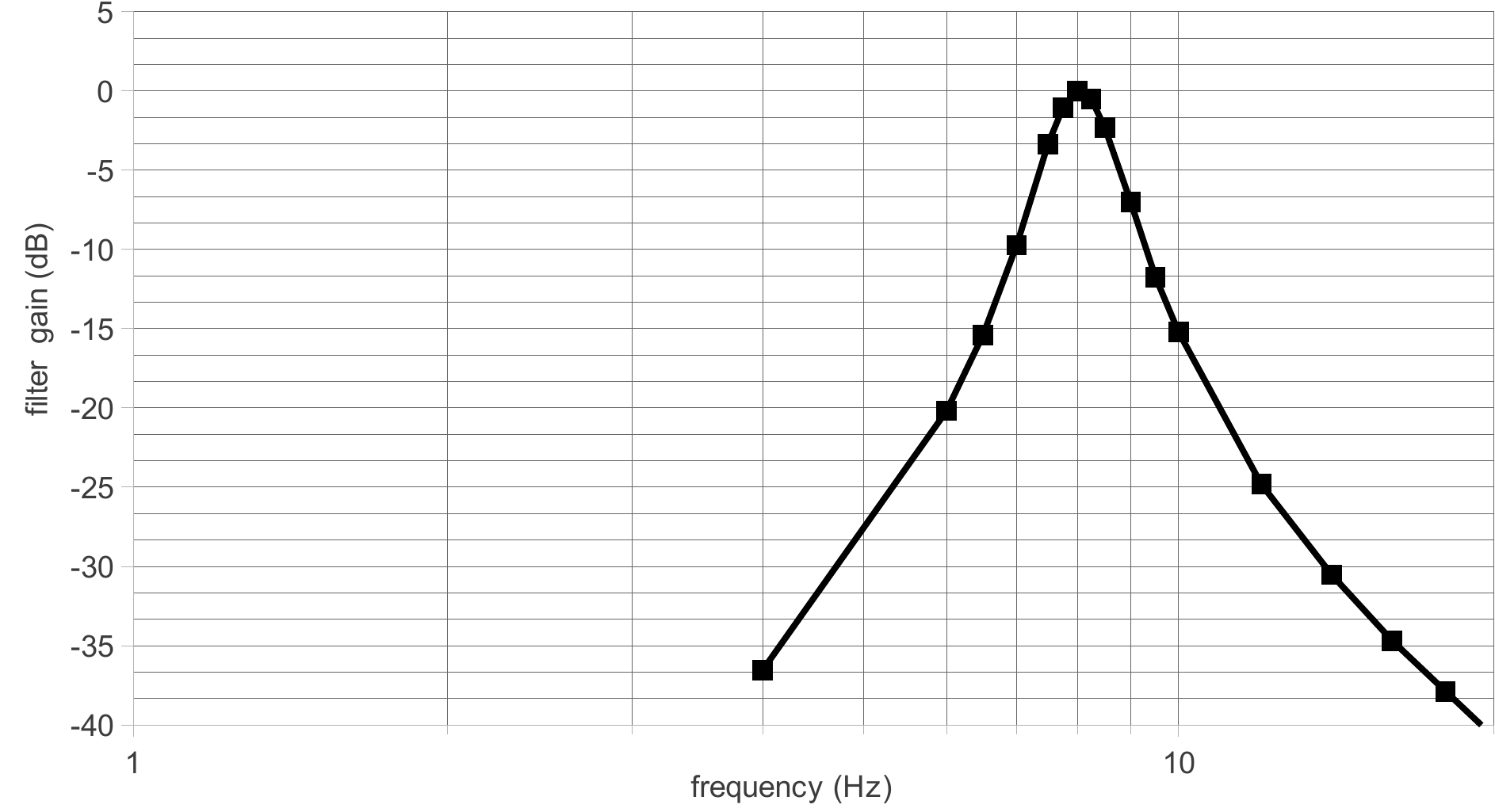}
\caption{PMT input bandpass filter frequency response measurement. The filter parameters are verified, i.e center frequency lying at 8\,Hz and a quality factor of 8.}
\label{reponseFiltre8Hz} 
\end{center}
\end{figure}

At first, the AC input band pass filter was characterized. To this end, one of the mercury `8\,Hz' outputs was connected to the PMT signal input, and its frequency was varied from 0.2\,Hz to 11.5\,Hz. 
The result of the measurement, displayed in \fig\ref{reponseFiltre8Hz}, is in accordance with a Q\,=\,8 filter, centered 
at 8\,Hz.

\begin{figure}[ht]
\begin{center}
\includegraphics[angle=0,width=0.95\textwidth]{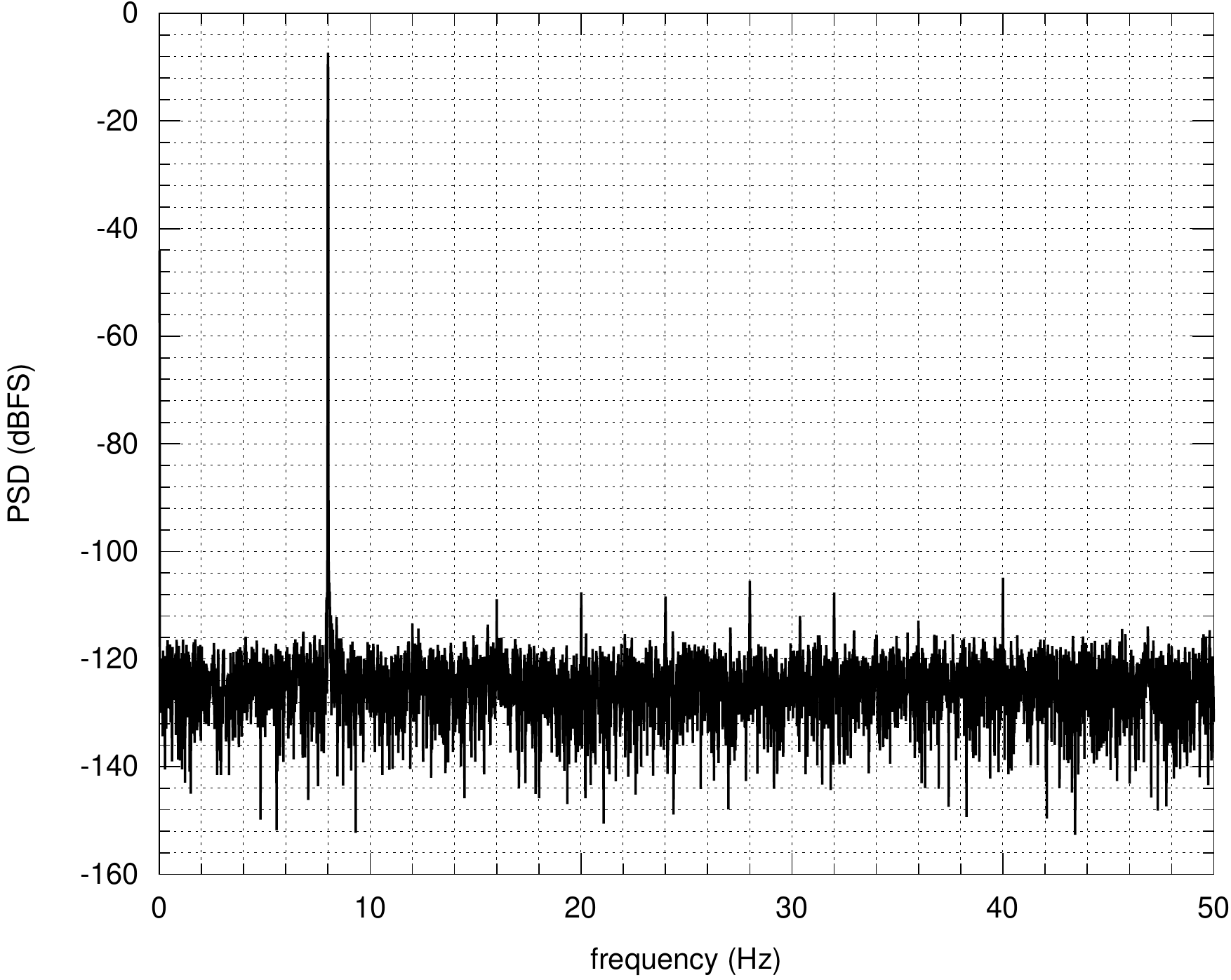}
\caption{FFT of the digitized input signal in loop-back mode. A sampling rate of 100\,Hz and an acquisition duration of a 100\,s were used.}
\label{DDS_8Hz_reboucle_100s_100Hz} 
\end{center}
\end{figure}

The next crucial test consisted in checking that the intrinsic board noise was negligible as compared to the physical noise which is obviously one of the parameter limiting the precision. The Power Spectrum Density (PSD) of the complete board input chain was measured: this includes the amplifier (tested with a gain 1), the band pass filter and the ADC. As in the previous measurement, the input signal was a mercury `8\,Hz' signal generated by the board itself. A sampling rate of 100\,Hz and an acquisition time of 100\,s were chosen as they represent typical values used by the experiment. The Fast Fourier Transform of the digitized signal is shown in \fig\ref{DDS_8Hz_reboucle_100s_100Hz}. The measured noise floor is compatible with the expected ADC performances as advertised by the manufacturer for the same amount of points. It is also worth noting that the noise floor inside and outside of the filter bandwidth is identical, indicating that the electronics noise is due to quantization only.
The signal-to-noise ratio obtained from the PSD spectrum is about 14000. 

\begin{figure}[ht]
\begin{center}
\includegraphics[angle=0,width=0.95\textwidth]{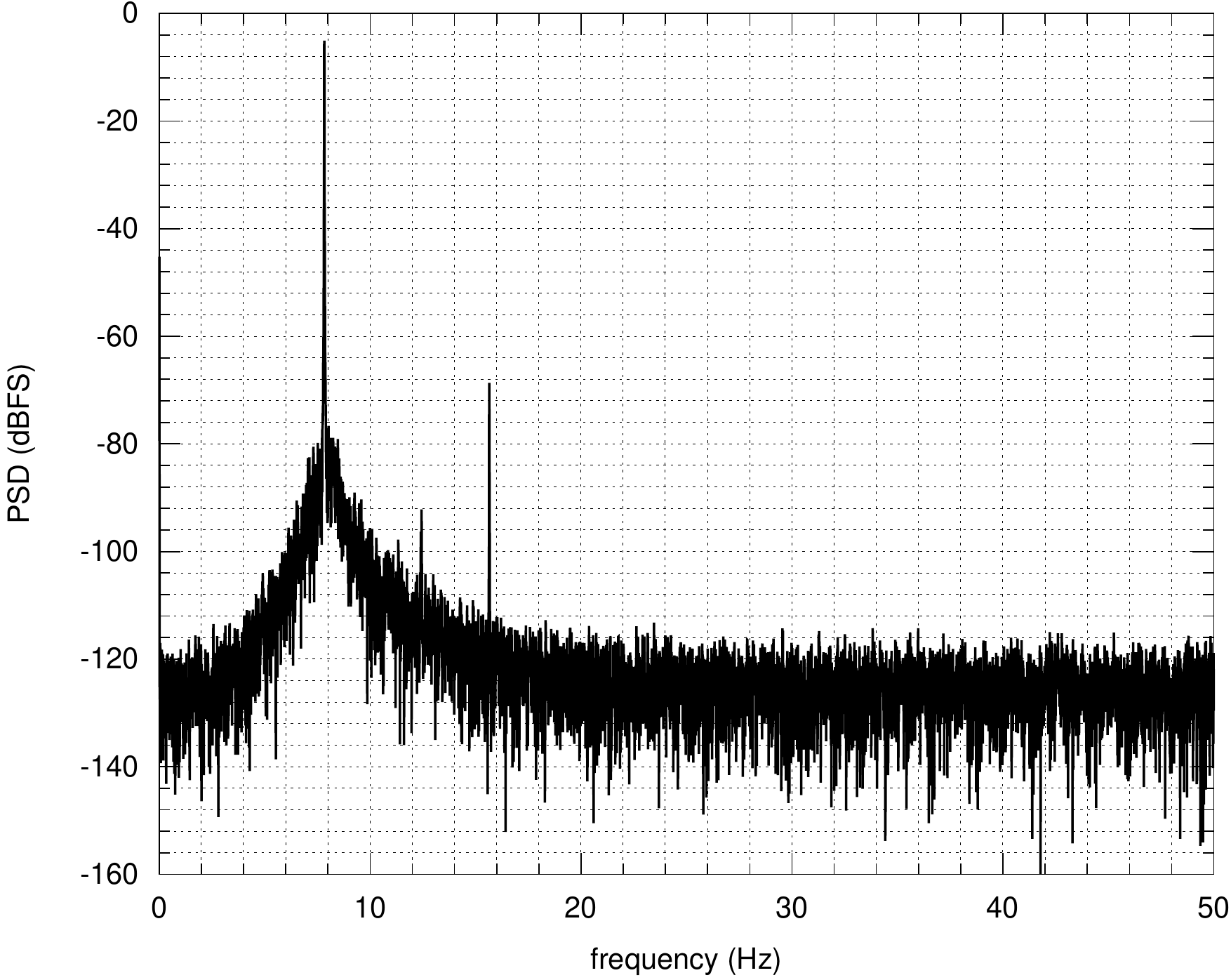}
\caption{FFT of the digitized precession signal. A sampling rate of 100\,Hz and an acquisition duration of a 100\,s were used.}
\label{PSD_PSI} 
\end{center}
\end{figure}

Those performances should be compared with the precession signal measurement in the experimental setup. The PSD of a typical measurement is shown in \fig\ref{PSD_PSI}. In contrast to the loop-back measurement, the noise floor greatly increases in the filter bandwidth, showing that the measurement noise is dominated by the physical noise.
 The signal-to-noise ratio obtained from the PSD spectrum is about 840.

From this test, we can conclude that the sensitivity in the frequency extraction of the mercury signal will not be limited by the electronic performances (security coefficient of 10).

\section{Summary}
\label{SummarySec}
We have designed and constructed an electronic board for the data acquisition and control of the neutron electric dipole moment experiment at the Paul Scherrer Institute. The board, organized around a Field Programmable Gate Array (FPGA), is a multifunction module which fulfills a large fraction of the requirements regarding the time-sequencing and the data acquisition of the experiment. Thanks to the use of a FPGA, it is compact, versatile and evolutive. Additionally, the high integration level and the corresponding limited number of connections should result in a better reliability.
A list of the main functions and specifications is given in table~\ref{perfTable}. It must be noted that all requirements given in section~\ref{requirements} are all easily met.

\begin{table}[th]
\begin{center}
\begin{tabular}{|l|l|}
\hline
   Function & Specifications \\
\hline
   \multirow{2}{*} {Microtimer}       &  16 time sequences with 32 actions  \\
                                      &  1\,\textmu s time resolution \\
\hline
   \multirow{5}{*} {Excitation pulse} &  frequency resolution 1.77\,nHz \\
                                      &  phase resolution 0.011\,\textdegree \\
                                      &  amplitude resolution 12 bit \\
                                      &  output current range $\pm20$\,mA \\
                                      &  ramping time:  32.768\,ms to 2.096\,s (8.192\,ms resolution)\\
\hline
   \multirow{3}{*} {Spin flipper}     &  frequency resolution 37\,mHz \\
                                      &  amplitude resolution 10 bit \\
                                      &  output voltage range $\pm5$\,V \\
\hline
   \multirow{6}{*} {PMT acquisition}     &  16 bit digitization for 3.3\,V \\
                                      &  input current conversion factor: 1\,V/\textmu A \\
                                      &  bandpass input filter  4\textsuperscript{th} order, $\rm f_c$=8, Q=8\\
                                      &  selectable voltage gain: 1 to 1000\\
                                      &  sampling period: 10\,\textmu s to 1677\,ms (100\,ns resolution) \\
                                      &  SNR = 14,000  at full scale\\
\hline
\multirow{3}{*} {Slow monitoring}     &  16 bit digitization \\
                                      &  input voltage range: 0-3.3\,V \\
                                      &  sampling rate of up to 50\,Hz\\
\hline
\multirow{2}{*} {Neutron counter}     &  maximum counting rate of 150\,MHz \\
                                      &  full scale 2\textsuperscript{32} \\
\hline
\end{tabular}
\end{center}
\caption{Summary of the module specifications.}
\label{perfTable}
\end{table}


\end{document}